\documentclass[secnumarabic,aps,prd,floatfix,nofootinbib,showkeys,twocolumn,showpacs,10pt,superscriptaddress] 
{revtex4-1}

\usepackage{times}
\usepackage{amsfonts}
\usepackage{amsmath}
\usepackage{amssymb}
\usepackage{natbib}
\usepackage{graphicx}
\usepackage{enumitem}
\usepackage{xcolor}
\usepackage[colorlinks=true,linkcolor=red, citecolor=blue]{hyperref}
\usepackage[outdir=./]{epstopdf}
\usepackage[normalem]{ulem}
\usepackage{amsmath}
\usepackage{booktabs}

\usepackage[utf8]{inputenc}
\usepackage{natbib}
\usepackage{hyperref}

\def\araa{ARAA}
\def\aap{A\&A}

\def\apj{ApJ}

\def\jcap{JCAP}
\def\mnras{MNRAS}
\def\na{New Astronomy}

\begin{document}
\definecolor{orange}{rgb}{0.9,0.45,0}
\def\CovDev{D}
\def\Res{{\mathcal R}}
\def\Gammaflat{\hat \Gamma}
\def\metricflat{\hat \gamma}
\def\Dflat{\hat {\mathcal D}}
\def\part_n{\partial_\perp}
%
\def\Lie{\mathcal{L}}
\def\A{\mathcal{X}}
\def\Aphi{\A_{\phi}}
\def\hAphi{\hat{\A}_{\phi}}
\def\E{\mathcal{E}}
\def\Ham{\mathcal{H}}
\def\M{\mathcal{M}}
\def\R{\mathcal{R}}
\def\p{\partial}
\def\hg{\hat{\gamma}}
\def\hA{\hat{A}}
\def\hD{\hat{D}}
\def\hE{\hat{E}}
\def\hR{\hat{R}}
\def\hcA{\hat{\mathcal{A}}}
\def\hDelt{\hat{\triangle}}
\def\na{\nabla}
\def\dif{{\rm{d}}}
\def\non{\nonumber}
\newcommand{\erf}{\textrm{erf}}
\newcommand{\saeed}[1]{\textcolor{blue}{SF: #1}} 
%
\renewcommand{\t}{\times}
\long\def\symbolfootnote[#1]#2{\begingroup%
\def\thefootnote{\fnsymbol{footnote}}\footnote[#1]{#2}\endgroup}
\title{Luminosity-Temperature Relation as a Probe for Modified Gravity} 

\author{Antonino Del Popolo}
\email{ antonino.delpopolo@unict.it}
\affiliation{Dipartimento di Fisica e Astronomia, Universit´a di Catania, Italia}
\affiliation{INFN sezione di Catania, Via S. Sofia 64, I-95123 Catania, Italy}

\author{Saeed Fakhry}
\email{saeed.fakhry@uv.es}
\affiliation{Departamento de Astronom\'{\i}a y Astrof\'{\i}sica, Universitat de Val\`encia, Dr. Moliner 50, 46100 Burjassot (Valencia), Spain}

\author{David F. Mota}
\email{d.f.mota@astro.uio.no}
\affiliation{Institute of Theoretical Astrophysics, University of Oslo, N-0315 Oslo, Norway}

\date{\today}

\begin{abstract}
We investigate the luminosity-temperature ($L$-$T$) relation of galaxy clusters as a probe for testing modified gravity (MG) theories, focusing on $f(R)$ gravity and symmetron models. Using an improved semi-analytic framework that incorporates angular momentum acquisition, dynamical friction, and shock heating within the modified punctuated equilibrium model, we compare predictions against hydrodynamical simulations and observational data. While massive clusters remain largely screened and follow standard $\Lambda$CDM predictions, low-mass systems ($kT \lesssim 1$-2 keV) exhibit systematic deviations characterized by steeper $L$-$T$ slopes in MG scenarios. Crucially, we demonstrate that these signatures cannot be mimicked by conventional astrophysical processes such as feedback or angular momentum effects, which primarily affect normalization rather than curvature. Our results establish the $L$-$T$ relation as a robust diagnostic tool for distinguishing general relativity from screened MG theories, with the strongest discriminatory power emerging at group scales accessible to current and future X-ray surveys. Moreover, a normalized reduced $\chi^2$ analysis of the $L$-$T$ relation shows that MG models provide significantly better agreement with observational data than $\Lambda$CDM, with several realizations achieving excellent fits while the $\Lambda$CDM model consistently performs worst.

\end{abstract}

\maketitle
\vspace{0.8cm}


\section{Introduction}
The standard $\Lambda$CDM cosmological model has achieved remarkable precision in describing the large-scale structure of the Universe and the statistical properties of the Cosmic Microwave Background (CMB) \cite{Planck:2020, Riess:2019}. However, the persistent nature of small-scale challenges, such as the ``cusp-core" and ``missing satellites" problems, alongside the growing $H_0$ and $S_8$ tensions, has fueled interest in modified gravity (MG) theories \cite{DiValentino:2021, Joyce:2016}. Models like $f(R)$ gravity \cite{Hu:2007} and symmetron scalar-tensor theories \cite{Hinterbichler:2010} provide a compelling alternative to general relativity (GR) by explaining cosmic acceleration through a ``fifth force" while employing screening mechanisms to remain consistent with Solar System constraints \cite{Khoury:2004, Brax:2012, DeFelice:2010}. Distinguishing these gravitational modifications from the standard paradigm remains one of the primary goals of modern observational cosmology.

Galaxy clusters, as the most massive virialized objects in the cosmic web, serve as critical probes for testing these competing theories \cite{Allen:2011, Kravtsov:2012}. The growth and thermal state of the Intra-Cluster Medium (ICM) are dictated by the underlying gravitational potential, which in turn reflects the nature of dark matter and the laws of gravity \cite{Giodini:2009, Borgani:2011}. Consequently, the scaling relations between fundamental cluster properties, such as mass ($M$), X-ray luminosity ($L$), and gas temperature ($T$), are expected to carry signatures of the gravitational framework \cite{Kaiser:1986, Arnaud:1999}. While self-similar models predict simple power-law relations, observations consistently reveal a ``break" or steepening, particularly at the low-mass end of the cluster population (galaxy groups), where non-gravitational physics becomes dominant \cite{Pratt:2009, Maughan:2012, Lovisari:2015}.

The utility of these scaling relations as a test for gravity has recently been called into question due to the role of astrophysical nuisances. It was shown by \cite{DelPopolo:2019} that the $M\mbox{-}T$ relation is a poor probe for distinguishing between $\Lambda$CDM and MG because an improved top-hat collapse model can effectively mimic the effects of screened gravity. By incorporating ordered angular momentum from tidal-torque interactions \cite{White:1984, DelPopolo:2009}, random angular momentum, dynamical friction \cite{ElZant:2001, DelPopolo:2013}, and modifications to the virial theorem including surface pressure \cite{Afshordi:2002}, the $M\mbox{-}T$ relation in $\Lambda$CDM exhibits a non-self-similar behavior and a break at $T \sim 3\mbox{-}4$ keV. This astrophysical ``mimicry" produces a signature that is statistically indistinguishable from the temperature enhancements predicted by $f(R)$ or symmetron theories \cite{Hammami:2017, Arnold:2014}.

While the $M\mbox{-}T$ relation is theoretically fundamental, mass is not a direct observable and is often subject to hydrostatic bias or lensing systematics \cite{Rasia:2006, Applegate:2014}. In contrast, the X-ray luminosity is a more directly accessible quantity, though it is highly sensitive to the gas density profile and the thermal history of the cluster \cite{Bryan:1998, Voit:2005}. The $L\mbox{-}T$ relation, in particular, has long been observed to be steeper than the self-similar expectation ($L \propto T^2$), an effect usually attributed to AGN feedback or pre-heating \cite{LeBrun:2014, Giodini:2013, Truong:2018}. If the underlying $M\mbox{-}T$ relation is already ``contaminated" by the acquisition of angular momentum and dynamical friction as suggested by \cite{DelPopolo:2019}, then any observable derived from mass, specifically luminosity, will likely inherit these degeneracies, complicating the use of $L\mbox{-}T$ as a probe for MG.

In this paper, we propose to study the $L\mbox{-}T$ relation as a probe for MG theories. The structure of this paper is as follows. In Sec.\,\ref{sec:ii}, we review the theoretical framework of $f(R)$ gravity and symmetron models, describing the scalar-tensor formalism and the hydrodynamical simulations used as benchmarks for our analysis. In Sec.\,\ref{sec:iii}, we develop an improved semi-analytic model for the $L\mbox{-}T$ relation, incorporating angular momentum acquisition, dynamical friction, and shock heating within the modified punctuated equilibrium framework. In Sec.\,\ref{sec:iv}, we present our main results, comparing predictions from $\Lambda$CDM and MG models against observational data from galaxy cluster surveys, with particular emphasis on the low-mass regime where screening mechanisms become inefficient. Finally, in Sec.\,\ref{sec:iv}, we summarize our findings and discuss the implications for constraining MG theories with current and future observations.

\section{Modified Gravity: Models and Simulations}\label{sec:ii}

In this section, we summarize the theoretical framework of the MG theories considered in this work, specifically focusing on the $f(R)$ gravity and symmetron models used in the hydrodynamical simulations of \cite{Hammami:2017}. These models belong to the class of scalar-tensor theories, where the gravitational interaction is mediated not only by the metric tensor $g_{\mu\nu}$ but also by a scalar field $\phi$. The general action describing these theories in the Einstein frame is given by:
\begin{equation}
    S = \int d^4x \sqrt{-g} \left[ \frac{1}{2}M_{pl}^2 R - \frac{1}{2} \partial^i \phi \partial_i \phi - V(\phi) \right] + S_m(\tilde{g}_{\mu\nu}, \psi_i),
\end{equation}
where $R$ is the Ricci scalar, $M_{pl}$ is the reduced Planck mass, and $V(\phi)$ represents the self-interacting potential of the scalar field. The matter fields $\psi_i$ are coupled to a Jordan frame metric $\tilde{g}_{\mu\nu}$, which is related to the Einstein frame metric via a conformal transformation $\tilde{g}_{\mu\nu} = A^2(\phi)g_{\mu\nu}$, where $A(\phi)$ is the conformal coupling function. This coupling gives rise to a fifth force acting on massive particles:
\begin{equation}
    \vec{F}_{\phi} = -\frac{A'(\phi)}{A(\phi)} \vec{\nabla}\phi,
\end{equation}
where the prime denotes a derivative with respect to the scalar field.

\subsection{The Symmetron Model}
The symmetron model \cite{Hinterbichler:2010} relies on a screening mechanism where the scalar field acquires a vacuum expectation value (VEV) in low-density regions, leading to a fifth force, while in high-density regions, the field is driven toward $\phi = 0$, restoring GR. This behavior is achieved through a potential and a coupling function of the form:
\begin{equation}
    V(\phi) = V_0 - \frac{1}{2}\mu^2\phi^2 + \frac{1}{4}\lambda\phi^4, \quad A(\phi) = 1 + \frac{1}{2} \left( \frac{\phi}{M} \right)^2,
\end{equation}
where $\mu$ and $M$ are mass scales and $\lambda$ is a dimensionless parameter. The fifth force in this model can be expressed in terms of the strength of the scalar field $\beta$ and the range of the force $\lambda_0$. In the supercomoving coordinates, the force reads:
\begin{equation}
    F_{\phi} = 6\Omega_m H_0^2 \frac{\beta^2 \lambda_0^2}{a_{SSB}^3} \tilde{\phi} \tilde{\nabla} \tilde{\phi},
\end{equation}
where $a_{SSB}$ is the expansion factor at the time of symmetry breaking. This force can significantly alter the virialization process of galaxy clusters, potentially increasing the gas temperature for a given halo mass \cite{Winther:2012}.

\subsection{$f(R)$ Gravity}
In $f(R)$ gravity, the Ricci scalar in the Einstein-Hilbert action is replaced by a non-linear function $f(R)$ \cite{DeFelice:2010, Sotiriou:2010}. We adopt the popular Hu-Sawicki model \cite{Hu:2007}, which is designed to reproduce the $\Lambda$CDM expansion history while evading Solar System tests via the chameleon screening mechanism. The functional form is:
\begin{equation}
    f(R) = -m^2 \frac{c_1(R/m^2)^n}{1 + c_2(R/m^2)^n},
\end{equation}
where $m^2 \equiv H_0^2 \Omega_m$. In the high-curvature regime ($R \gg m^2$), the model parameters are often simplified such that $c_1/c_2 \approx 6\Omega_\Lambda / \Omega_m$. The model is typically characterized by the value of the derivative $f_{R0} = df/dR$ at the current epoch. Small values (e.g., $|f_{R0}| = 10^{-6}$) represent weak modifications, while larger values (e.g., $10^{-4}$) lead to significant deviations from GR. For the purpose of force calculation, this is equivalent to a scalar-tensor theory with a coupling $\beta = \sqrt{1/6}$ and a fifth force:
\begin{equation}
    F_{\phi} = -\frac{a}{2\beta M_{pl}} \nabla \phi.
\end{equation}
These modifications lead to an enhancement of the gravitational potential, which increases the depth of the cluster's potential well and results in higher X-ray luminosities and temperatures compared to $\Lambda$CDM for a fixed mass \cite{Arnold:2014}.

\subsection{Simulations and the L-T Relation}
To evaluate the impact of these theories on the $L\mbox{-}T$ relation, we compare our semi-analytic model results with the hydrodynamical N-body simulations performed by \cite{Hammami:2017} using the ISIS code \cite{Llinares:2014}. These simulations include both $f(R)$ and symmetron models, with box sizes of 200 Mpc/h and 256 Mpc/h respectively. Crucially, these simulations provide the $M\mbox{-}T$ and $L\mbox{-}T$ baselines that we use to calibrate our improved top-hat model. Following the logic of \cite{DelPopolo:2019}, we incorporate the mass-dependent effects of angular momentum acquisition and dynamical friction into the calculation of $T$ and then transform these to $L$ using the $M/L$ scaling relations found in \cite{Mantz:2010, Popesso:2005}. This approach allows us to quantify the degree to which astrophysical processes in $\Lambda$CDM can overlap with the signals of MG in the $L\mbox{-}T$ plane \cite{Borgani:2004, Stanek:2010}.

\section{Theoretical Model}\label{sec:iii}

\subsection{The Mass-Temperature Relation}
The classical self-similar scaling, $M \propto T^{3/2}$, arises from the assumption that the only scale in the problem is the virial radius $r_{vir}$ \cite{Kaiser:1986, Bryan:1998}. To improve this, we consider the radial acceleration of a shell of matter, accounting for the ordered angular momentum acquired via tidal-torque interactions with neighboring protostructures \cite{White:1984, Catelan:1996}, random angular momentum, and dynamical friction $\eta$ \cite{Chandrasekhar:1943, DelPopolo:2013}. The radial equation of motion for a mass element in the presence of a cosmological constant $\Lambda$ is given by:
\begin{equation}
\frac{d v_r}{dt}
=
-\frac{G M}{r^2}
+\frac{L^2}{M^2 r^3}
+\frac{\Lambda}{3} r
-\eta\, v_r ,
\label{eq:motion}
\end{equation}
where $L$ denotes the total (ordered plus random) angular momentum, $\Lambda$ is the cosmological constant, and $\eta$ is the dynamical friction coefficient.

Integrating Eq.~(\ref{eq:motion}) yields the energy equation
\begin{equation}
\frac{1}{2}\left(\frac{dr}{dt}\right)^2
=
\frac{GM}{r}
+\int_0^r \frac{L^2}{M^2 r'^3} \, dr'
+\frac{\Lambda}{6} r^2
-\int_0^r \eta\,\frac{dr'}{dt}\,dr'
+\epsilon ,
\end{equation}
where $\epsilon$ is the binding energy of the shell, calculated from the turn-around condition $dr/dt=0$.

The total specific energy can be written as \cite{Voit:2000}
\begin{equation}
\epsilon =
-\frac{1}{2}
\left(\frac{2\pi G M}{t_\Omega}\right)^{2/3}
\left[
\left(\frac{M_0}{M}\right)^{5/(3m)}
-1
\right]
g(M),
\end{equation}
where $t_\Omega=\pi\Omega_{m,0}/[H_0(1-\Omega_{m,0})^{3/2}]$ is the characteristic collapse time, $m$ depends on the effective spectral index of perturbations, and the function $g(M)$ encapsulates deviations from self-similarity. The latter is given by
\begin{equation}
g(M)=
1+
\frac{F}{x-1}
+\frac{\lambda_0}{1-\mu(\delta)}
+\frac{\Lambda}{3H_0^2}\xi^3 ,
\end{equation}
with $H_0$ is the Hubble parameter at the present-time, $x=1+(t_\Omega/t)^{2/3}$, $M=M_0 x^{-3m/5}$, $\xi=r_{ta}/x_1$,where $r_{ta}$ is the turn-around radius,and $x_1$ is specified by $M=4\pi \rho_b x_1^{3}/3$, where $\rho_b$ shows the background density. Also, $F$ describes the contribution of angular momentum( see Eq.\,(6) of \cite{DelPopolo:2019}), and $\lambda_0=\epsilon_0 T_{c0}$ and $\mu(\delta)$ are given in \cite{1995ApJ...455...32C}.

The kinetic energy per unit mass, $E/M$, can then be related to the temperature through
\begin{equation}
k_B T
=
\frac{4}{3}\,
\tilde{a}
\frac{\mu m_p}{2\beta}
\frac{E}{M},
\end{equation}
also considered as the mass-temperature relation. In this relation, $\tilde{a}$ represents the ratio between kinetic and total energy, and $\beta$ relates the gas temperature to the dark matter velocity dispersion \cite{Voit:2000}.

Following a series of intermediate steps, the improved mass–temperature relation takes the form:
\begin{equation}
k_B T \propto M^{2/3}\frac{m(M)}{n(M)},
\end{equation}
where
\begin{equation}
m(M) = \frac{1}{m} + \left(\frac{t_{\Omega}}{t}\right)^{2/3} + \frac{K(m, x)}{(M/M_0)^{8/3}} + \frac{\lambda_0}{1 - \mu(\delta)} + \frac{\Lambda\xi^3}{3H_0^2 \Omega_{m,0}},
\end{equation}
and
\begin{equation}
n(M) = \frac{1}{m} + \left(\frac{t_{\Omega}}{t_0}\right)^{2/3} + K_0(m, x).
\end{equation}
In the above relations, $K(m,x)$ can be represented using the LerchPhi function \cite{DelPopolo:2009}.
\subsection{The Luminosity-Temperature Relation}
Within the framework of the modified punctuated equilibrium model (MPEM) \cite{Cavaliere1998f, DelPopolo:2019}, the thermodynamical state of the intracluster medium (ICM) is determined by a sequence of merger-driven shock-heating events interspersed with quasi-hydrostatic phases. In this picture, the X-ray luminosity–temperature ($L$-$T$) relation naturally emerges from the interplay between gravitational heating, accretion history, and the internal structure of galaxy clusters.

The bolometric X-ray luminosity generated by thermal bremsstrahlung emission is given by
\begin{equation}
L = \frac{6\pi k}{C_1(\mu m_p)^2} \int_0^{R_{\rm vir}} r^2 \rho_g^2(r) T_g^{1/2}(r) dr ,
\label{eq:lum_basic}
\end{equation}
where $\mu=0.59$ is the mean molecular weight, $C_1$ is the bremsstrahlung normalization constant, and $\rho_g(r)$ and $T_g(r)$ denote the gas density and temperature profiles, respectively. In normalized form, this becomes \cite{1998ApJ...501..493C}
\begin{equation}
L \propto \int_0^{r_2} n^2(r) T^{1/2}(r) d^3r ,
\label{eq:lum_norm}
\end{equation}
where $T(R)$ is the temperature in the plasma, and $r_2$ denotes the shock radius, approximately coincident with the virial radius.

In the MPEM framework, accreting gas is shock-heated at $r\simeq r_2$. The post-shock temperature is determined by the Rankine–Hugoniot conditions, with the pre-shock temperature given by
\begin{equation}
kT_1 = \max \left( kT_{1v}, kT_{1*} \right),
\end{equation}
where $kT_{1v}$ is the virial temperature of the infalling substructure and $kT_{1*}\simeq0.5,\mathrm{keV}$ accounts for preheating. The post-shock temperature is then
\begin{equation}
kT_2 = \frac{\mu m_p v_1^2}{3}\left[\frac{(1+\sqrt{1+\epsilon})^2}{4} + \frac{7}{10}\epsilon - \frac{3}{20}\frac{\epsilon^2}{(1+\sqrt{1+\epsilon})^2}\right],
\end{equation}
with $\epsilon = 15kT_1/(4\mu m_p v_1^2)$ and $v_1$ the infall velocity set by the gravitational potential at $r_2$.

The corresponding density jump across the shock is
\begin{equation}
\frac{n_2}{n_1} = 2\left(1-\frac{T_1}{T_2}\right) +
\left[4\left(1-\frac{T_1}{T_2}\right)^2 + \frac{T_1}{T_2}\right]^{1/2}.
\end{equation}

Assuming a polytropic equation of state, $T(r)/T_2=[n(r)/n_2]^{\gamma-1}$ with $\gamma\simeq1.2$, and hydrostatic equilibrium, the luminosity can be expressed as
\begin{eqnarray}
L \propto \left(\frac{n_2}{n_1}\right)^2\rho
\left(\frac{T_2}{T_v}\right)^{1/2}
\overline{\left[\frac{n(r)}{n_2}\right]^{2+(\gamma-1)/2}}
 M^{4/3} \times \nonumber \\
 \sqrt{\frac{\frac{1}{m}+\left(\frac{t_\Omega}{t}\right)^{2/3}+\frac{K}{\left(\frac{M}{M_0}\right)^{8/3}}+\frac{\lambda_0}{1-\mu(\delta)}\frac{\Lambda\xi^{3}}{3H_{0}^2\Omega_{m, 0}}}{\frac{1}{m}+\left(\frac{t_\Omega}{t}\right)^{2/3}+\frac{K_0}{M_0^{8/3}}}}
\label{eq:L_M}
\end{eqnarray}
where $T_v$ is the virial temperature and the overbar denotes a volume average.

Using the improved mass–temperature relation, the luminosity can be written as
\begin{equation}
L(T) \propto \mathcal{A}(M) T^2 \left[\frac{n(M)}{m(M)}\right]^{3/2},
\label{eq:LT_final}
\end{equation}
where $\mathcal{A}(M)$ collects all mass-dependent contributions from the shock physics and density structure.

The key departure from self-similarity arises because the ratio $n(M)/m(M)$ is not constant. It depends on angular momentum acquisition, dynamical friction, and cosmological terms entering the collapse dynamics:
\begin{equation}
\begin{aligned}
\frac{n(M)}{m(M)} =
\frac{1/m + (t_\Omega/t_0)^{2/3} + K_0}
{1/m + (t_\Omega/t)^{2/3} + K/(M/M_0)^{8/3}} \\
\qquad\qquad + \frac{\lambda_0}{1-\mu} + \frac{\Lambda \xi^3}{3H_0^2\Omega_{m,0}} \, .
\end{aligned}
\end{equation}
At high masses, this ratio approaches unity, recovering the self-similar scaling $L\propto T^2$. At group scales, however, angular momentum and dynamical friction terms dominate, causing $n(M)/m(M)$ to decrease and steepening the relation to $L\propto T^\alpha$ with $\alpha>2$, consistent with observed slopes $\alpha\simeq 3$–$3.5$.

The observed $L$-$T$ relation corresponds to an ensemble average over merger histories,
\begin{eqnarray}
\langle L \rangle = Q \int_{t-\Delta t}^{t} dt' \int_{0}^{m} dM' \int_{0}^{M-M'} d\Delta M \hspace{1cm}\nonumber\\
 \frac{df}{dM'}(M', t'|M, t)
\frac{d^2 p(M' \to M' + \Delta M)}{d\Delta M dt'} L,
\end{eqnarray}
with the associated variance obtained analogously
\begin{eqnarray}
\langle \Delta L^2 \rangle = Q \int_{t-\Delta t}^{t} dt'\int_{0}^{M} dM' \int_{0}^{M-M'} d\Delta M \hspace{1cm}\nonumber\\
\frac{df}{dM'}(M', t'|M, t) 
\frac{d^2 p(M' \to M' + \Delta M)}{d\Delta M dt'} (L - \langle L \rangle)^2.
\end{eqnarray}
This naturally introduces intrinsic scatter, which is larger for low-mass systems due to the wider diversity of accretion histories.

In MG theories, such as $f(R)$ or symmetron models, the effective gravitational coupling enhances the infall velocity and modifies the virial balance. This leads to (i) higher post-shock temperatures, (ii) altered density profiles, and (iii) modified collapse thresholds. Therefore, one would anticipate that high-mass clusters stay mostly screened and closely follow GR predictions, whereas smaller systems undergo more pronounced departures from standard behavior, resulting in a sharper slope in the $L\mbox{-}T$ relation at energies $kT\lesssim 1$–$2$ keV. This characteristic behaviour provides a sensitive observational discriminator between GR and MG scenarios.

\section{Results and Discussion}\label{sec:iv}

\begin{figure}[!ht]
 \centering
 \includegraphics[angle=0,width=0.95\hsize]{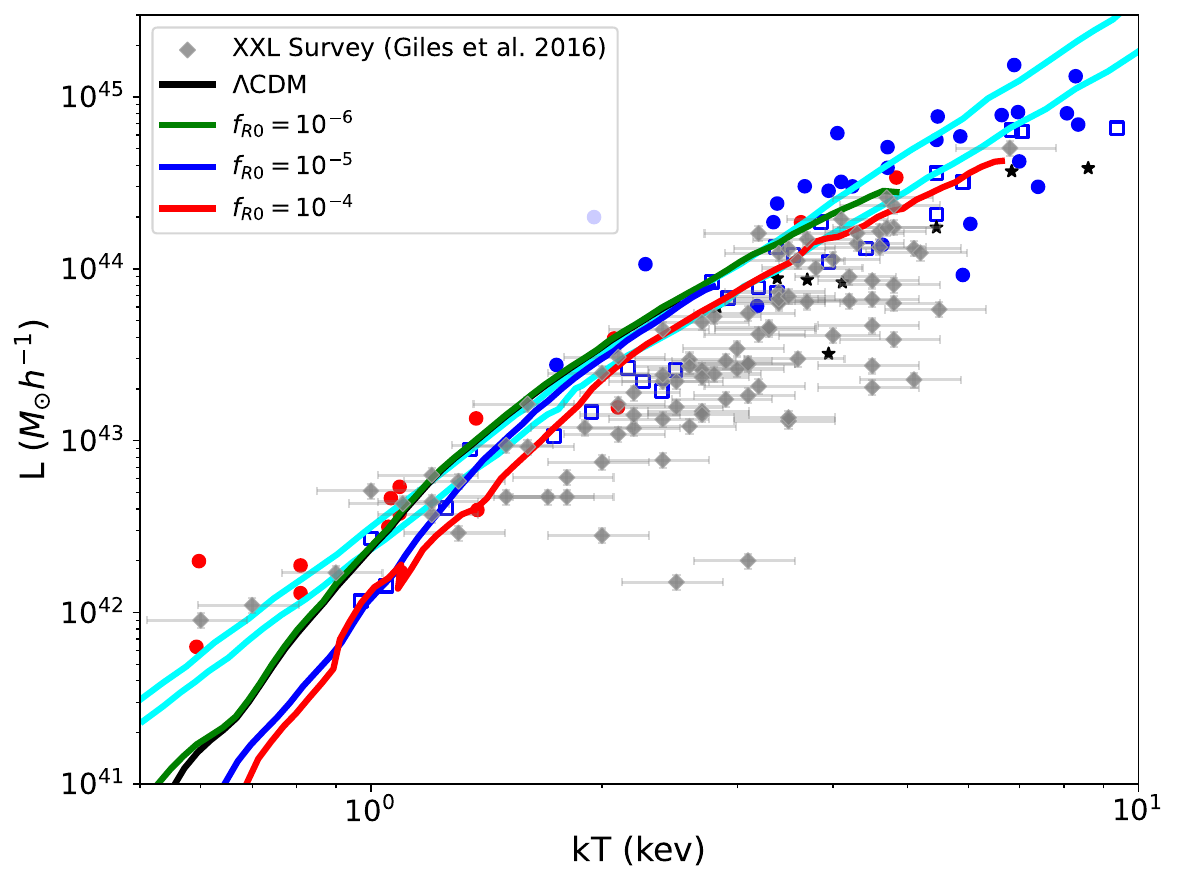}
\caption{The luminosity-temperature ($L$-$T$) relation for the $f(R)$ gravity models compared with the standard $\Lambda$CDM scenario. The black curve represents the $\Lambda$CDM prediction, while the stacked galaxy clusters are shown by red and blue circles, blue squares, and black stars. Also the corresponding data from XXL survey is presented as cyan points \cite{2016A&A...592A...3G}. The cyan shaded region denotes the 68\% confidence interval obtained from the continuous formation model [Eq.\,\eqref{eq:LT_final}] and the reference model of \citep{2002ApJ...564..669A}. The red, blue, and green curves correspond to the $f(R)$ models with three different parameters $f_{R0}=10^{-4}$, $10^{-5}$, and $10^{-6}$, respectively.}
\label{fig1}
\end{figure}

\begin{figure}[!ht]
 \centering
 \includegraphics[angle=0,width=0.95\hsize]{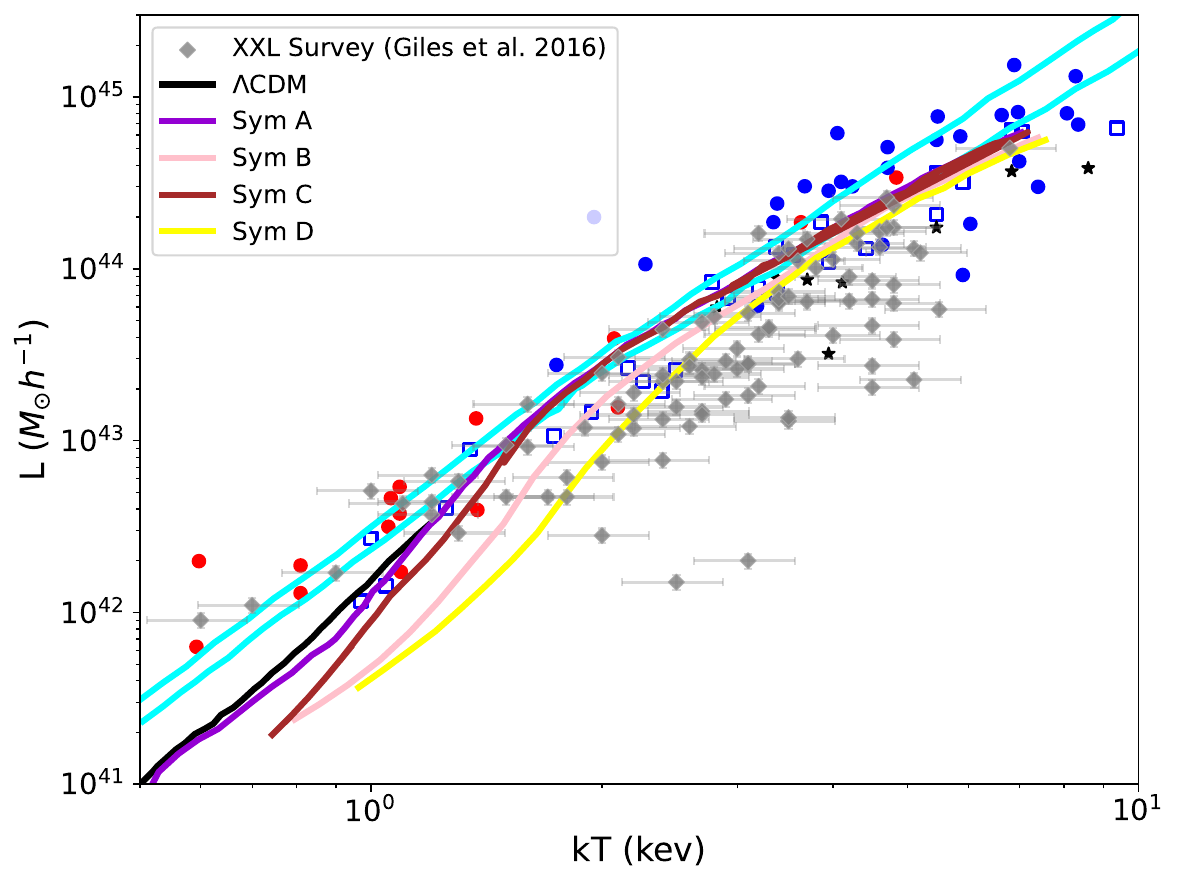}
\caption{Similar to Fig.\,\ref{fig1}, but for symmetron model. The black curve denotes the $\Lambda$CDM prediction, whereas the red, pink, brown, and yellow curves correspond to the symmetron models SymA, SymB, SymC, and SymD, respectively.}
\label{fig2}
\end{figure}

Figures \ref{fig1} and \ref{fig2} present the $L$-$T$ relation predicted by our theoretical framework and compare it with a wide compilation of observational data from galaxy cluster surveys, including the XXL sample as well as additional datasets covering both group- and cluster-scale systems. These observations span a broad range of masses and temperatures, allowing for a robust test of the underlying gravitational physics across different regimes.

In Fig.\,\ref{fig1}, we have shown the $L$-$T$ relation obtained in the context of $f(R)$ gravity, together with the corresponding $\Lambda$CDM prediction. The black solid curve represents the standard $\Lambda$CDM model, while the colored curves correspond to different values of the $f(R)$ parameter $|f_{R0}|=10^{-4}$, $10^{-5}$, and $10^{-6}$. The observational data points, drawn from multiple X-ray cluster samples, populate the region from rich clusters down to low-mass groups. A key feature emerging from the figure is that, although all models converge at the high-temperature end ($kT \gtrsim 3$-4 keV), significant deviations appear at lower temperatures, i.e., at $kT \lesssim 1$ keV. In particular, the $f(R)$ models predict a systematically steeper decline of the luminosity toward the group regime, in clear contrast with the $\Lambda$CDM expectation. Specifically, larger values of the scalaron field parameter $|f_{R0}|$ lead to increasingly pronounced deviations from the $\Lambda$CDM prediction, while smaller values correspond to progressively weaker departures.

This behaviour reflects the fact that, while massive clusters remain efficiently screened and therefore largely insensitive to modifications of gravity, lower-mass systems experience an enhanced effective gravitational force. As a consequence, the gas infall velocity, post-shock temperature, and density structure are modified, leading to a reduction of the X-ray luminosity at fixed temperature. Our results demonstrate that this effect cannot be mimicked by the standard astrophysical processes included in the $\Lambda$CDM framework, such as angular momentum acquisition or dynamical friction, which mainly affect the normalization but not the curvature of the $L$-$T$ relation.

A similar but even more pronounced behavior is observed in Fig.\,\ref{fig2} for the symmetron models. Here, the deviation from $\Lambda$CDM sets in at slightly higher temperatures, depending on the specific choice of the symmetry-breaking scale and coupling strength. The symmetron curves exhibit a sharper downturn at $kT \lesssim 1$-2 keV, reflecting the activation of the fifth force in low-density environments. Among the different realizations, models with stronger coupling or lower symmetry-breaking thresholds display the largest departure from the standard relation, reinforcing the sensitivity of the $L$-$T$ relation to the underlying scalar-field dynamics.

Importantly, the comparison with observational data shows that the low-temperature regime provides the strongest discriminatory power between gravity models. While the high-temperature end remains largely degenerate due to screening effects, the systematic offset observed at group scales constitutes a robust signature of MG. In this sense, our results demonstrate that the luminosity-temperature relation, unlike the mass–temperature relation, retains clear sensitivity to deviations from GR once realistic baryonic physics is included.

Our findings highlight the $L$-$T$ relation as a powerful and complementary probe of MG. The combination of its sensitivity to both gravitational dynamics and baryonic processes makes it particularly well suited for testing screened gravity models with current and forthcoming X-ray cluster surveys.

\subsection{Normalized Reduced $\chi^2$ Analysis}

We perform a quantitative statistical comparison between the observed $L$-$T$ relation of galaxy clusters and theoretical predictions from $\Lambda$CDM, $f(R)$ gravity, and Symmetron models. The analysis is based on a combined observational sample of clusters, including multiple subsamples with heterogeneous uncertainties.

For each theoretical model, the predicted luminosity is obtained by interpolating the model $L$-$T$ curve in logarithmic space at the observed cluster temperatures. The reduced $\chi^2$ statistic is then computed as
\begin{equation}
\tilde{\chi}^2 = \sum_{i=1}^{N}
\frac{\left(L_i^{\mathrm{obs}} - L_i^{\mathrm{model}}\right)^2}
{\sigma_{\mathrm{total},i}^2},
\end{equation}
where $L_i^{\mathrm{obs}}$ denotes the observed luminosity and $L_i^{\mathrm{model}}$ is the interpolated theoretical prediction. The total uncertainty $\sigma_{\mathrm{total},i}$ incorporates both luminosity and temperature errors, with the latter propagated into luminosity space using the local logarithmic slope of the model $L$-$T$ relation.

To facilitate a direct and scale-independent comparison between different gravity models, we define a normalized $\tilde{\chi}^2$ metric,
\begin{equation}
\tilde{\chi}^2_{\mathrm{norm}} =
\frac{\tilde{\chi}^2/(N-k)}{\max\!\left[\tilde{\chi}^2/(N-k)\right]},
\end{equation}
where $k=2$ approximates the effective number of free parameters in the scaling relation. This normalization rescales the $\tilde{\chi}^2$ values to the interval $[0,1]$, with lower values indicating better agreement with the data and $\tilde{\chi}^2_{\mathrm{norm}}=1$ corresponding to the worst-fitting model among those considered.

This framework enables a robust, quantitative ranking of $\Lambda$CDM, $f(R)$, and Symmetron models based on their ability to reproduce the observed cluster $L$-$T$ relation, beyond a purely visual comparison.
\begin{figure}[!ht]
 \centering
 \includegraphics[angle=0,width=0.95\hsize]{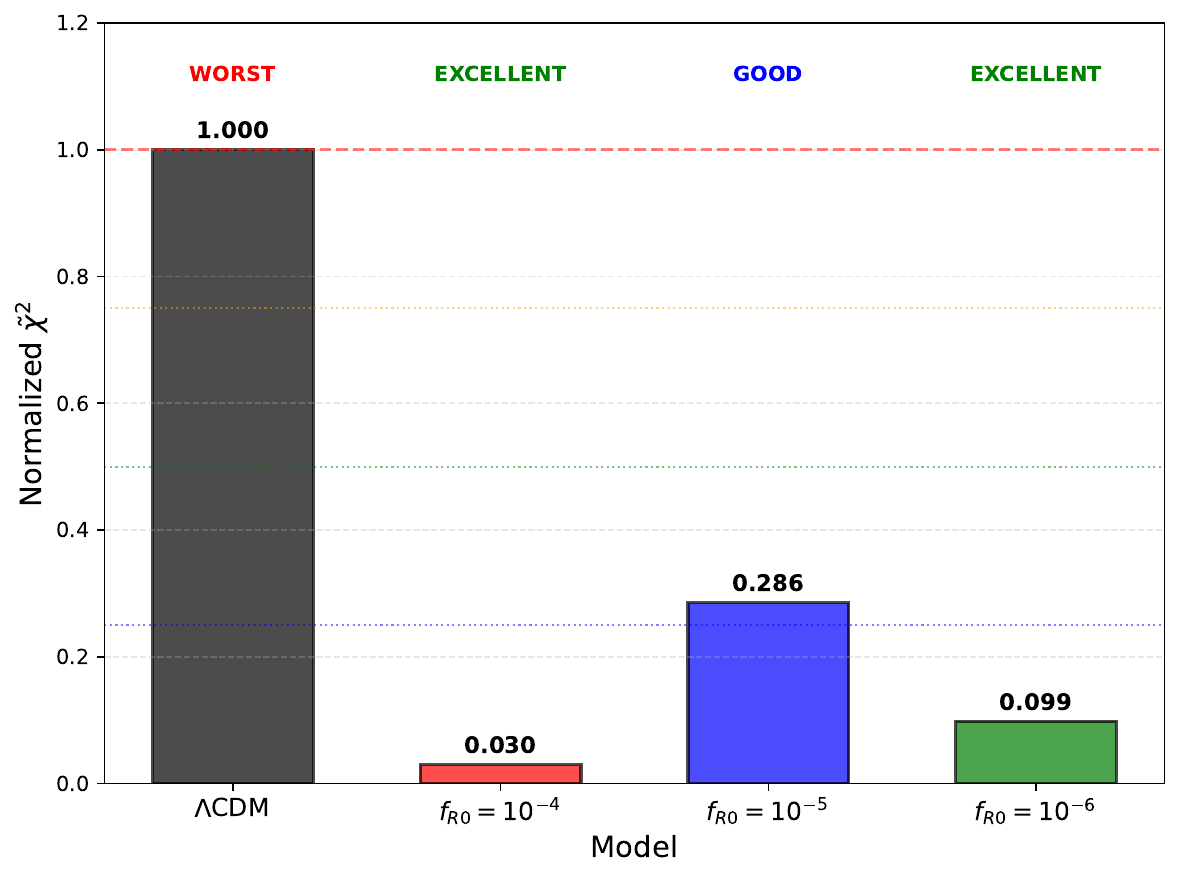}
\caption{Normalized $\tilde{\chi}^2$ analysis for the $f(R)$ gravity models, while comparing with the corresponding result from $\Lambda$CDM model.}
\label{fig3}
\end{figure}

\begin{figure}[!ht]
 \centering
 \includegraphics[angle=0,width=0.95\hsize]{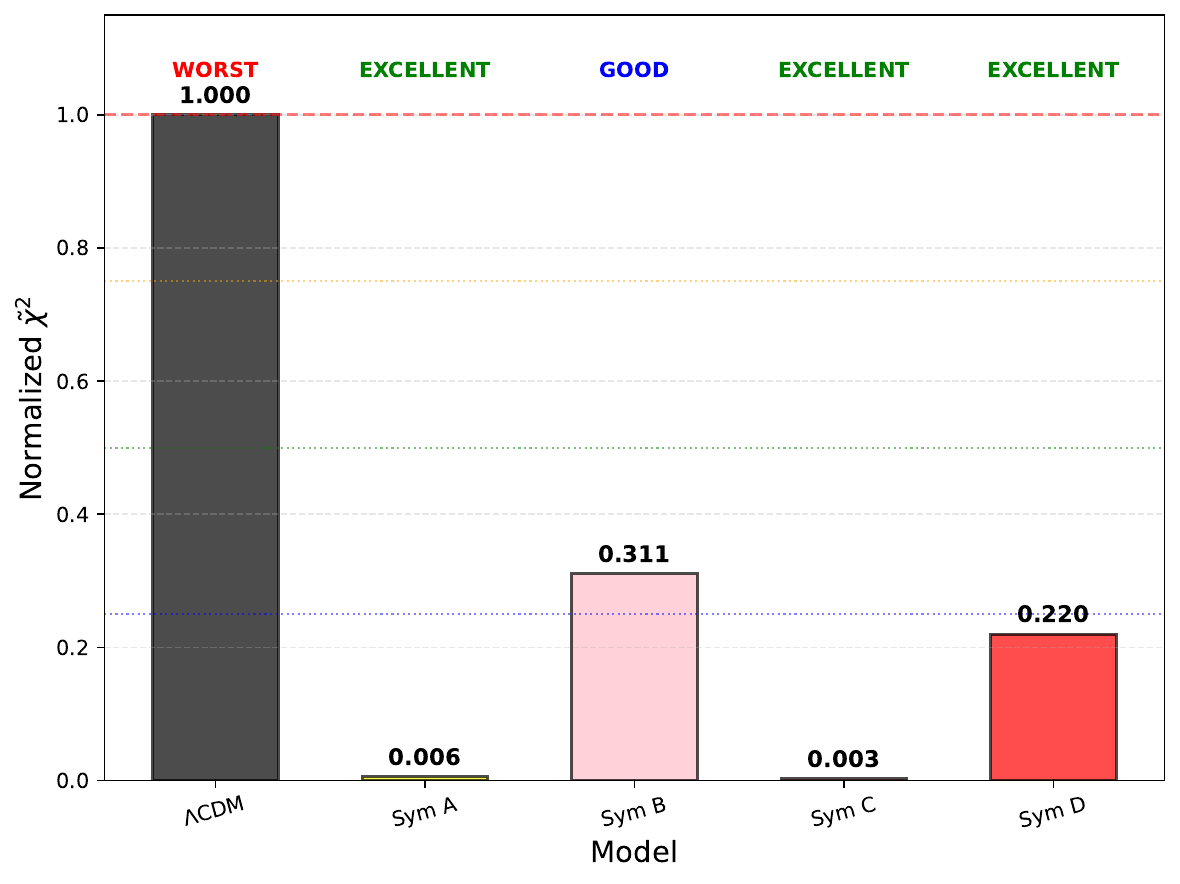}
\caption{Similar to Fig.\,\ref{fig3}, but for symmetron models.}
\label{fig4}
\end{figure}

In Fig.\,\ref{fig3}, we have depicted the normalized $\tilde{\chi}^2$ comparison for the $\Lambda$CDM and $f(R)$ models, providing a quantitative validation of the trends already identified in Fig.\,\ref{fig1}. While $\Lambda$CDM yields the worst agreement with the luminosity-temperature data, all $f(R)$ scenarios lead to a statistically significant improvement. In particular, the models with $|f_{R0}|=10^{-4}$ and $10^{-6}$ exhibit an excellent match to the observations, achieving the lowest normalized $\tilde{\chi}^2$ values, whereas the intermediate case $|f_{R0}|=10^{-5}$ still provides a good fit but with a slightly reduced performance. This hierarchy directly mirrors the visual behavior seen in Fig.\,\ref{fig1}, where MG effects primarily manifest at the low-temperature end, enhancing the curvature of the $L$-$T$ relation relative to $\Lambda$CDM. The consistency between the visual and statistical analyses demonstrates that the improved agreement of $f(R)$ gravity is not qualitative but quantitatively robust.

Moreover, in Fig.\,\ref{fig4}, we have extended the normalized $\tilde{\chi}^2$ analysis to the symmetron family of models and reinforces the conclusions drawn from Fig.\,\ref{fig2}. As in the $f(R)$ case, the $\Lambda$CDM model again provides the worst fit to the observational data, confirming its limited ability to reproduce the low-temperature cluster regime. Among the symmetron realizations, Sym~A, Sym~C, and Sym~D achieve an excellent level of agreement, yielding the smallest normalized $\tilde{\chi}^2$ values, while Sym~B remains good but systematically less successful than the best-performing variants. This ordering closely follows the degree to which each model enhances the deviation from self-similarity in the group-scale regime, exactly as observed in  Fig.\,\ref{fig2}. The strong coherence between the statistical ranking in  Fig.\,\ref{fig4} and the physical trends seen in the $L$-$T$ relation confirms that symmetron models can outperform $\Lambda$CDM in a controlled and predictive manner, rather than through fine-tuning or normalization shifts.

\section*{Conclusions}\label{sec:v}

In this work, we have investigated the $L$-$T$ relation of galaxy clusters as a diagnostic tool for testing modified theories of gravity, with particular emphasis on $f(R)$ gravity and symmetron models. By combining a semi-analytic framework based on the improved punctuated equilibrium model with results from state-of-the-art hydrodynamical simulations, we have explored how deviations fromGR manifest themselves in the thermodynamical properties of the intracluster medium. Our analysis demonstrates that the $L$-$T$ relation provides a sensitive and complementary probe to traditional mass-based observables, especially in regimes where screening mechanisms become inefficient.

A central result of this study is that MG models systematically predict departures from the standard $\Lambda$CDM $L$-$T$ relation at the low-temperature, group-scale regime. While massive clusters remain largely screened and thus follow a relation close to the standard self-similar expectation, low-mass systems exhibit a marked suppression of X-ray luminosity at fixed temperature. This behavior arises from the enhanced effective gravitational coupling in unscreened environments, which alters the collapse dynamics, shock heating efficiency, and gas density structure. In particular, both $f(R)$ and symmetron models predict a steeper $L$-$T$ relation below $kT \sim 1$-$2$ keV, in qualitative agreement with the trends seen in the numerical simulations.

A key outcome of our analysis is the demonstration that the deviations induced by MG cannot be trivially mimicked by conventional astrophysical processes. While angular momentum acquisition, dynamical friction, and non-gravitational heating processes significantly affect the normalization and scatter of the scaling relations, they do not reproduce the systematic curvature in the $L$-$T$ plane induced by MG. This result highlights a fundamental difference between baryonic feedback effects and genuine modifications of the underlying gravitational law, reinforcing the diagnostic power of the $L$-$T$ relation when interpreted within a physically motivated theoretical framework.

Our results also clarify the complementary roles of the $M$-$T$ and $L$-$T$ relations. While the $M$-$T$ relation is strongly affected by degeneracies between gravitational physics and baryonic processes, the $L$-$T$ relation retains a clearer imprint of MG, particularly in the low-mass regime. This makes the $L$-$T$ relation a more robust observable for distinguishing between $\Lambda$CDM and alternative theories, provided that high-quality observational data and accurate modeling of intracluster gas physics are available.

The normalized $\tilde{\chi}^2$ analysis plays a central role in our statistical assessment of the luminosity--temperature relation, enabling a direct and unbiased comparison between all gravity models considered. By rescaling the reduced $\chi^2$ values to a common reference, this approach highlights relative performance independently of absolute normalization or dataset size. The results show that screened MG models consistently provide superior agreement with observational data, with several realizations achieving excellent fits and others remaining statistically good, while the $\Lambda$CDM scenario systematically exhibits the poorest performance. This outcome confirms that the deviations introduced by MG are not marginal but statistically meaningful, and that they persist across the full temperature range probed by the data. Overall, the normalized $\tilde{\chi}^2$ metric strengthens the conclusion that MG offers a viable and potentially preferred framework for describing the thermodynamic scaling relations of galaxy clusters.

From an observational perspective, the comparison with current X-ray cluster samples indicates that the strongest discriminating power arises from galaxy groups and low-temperature clusters. Upcoming and ongoing surveys with improved sensitivity and spatial resolution, such as eROSITA and future X-ray missions, are therefore expected to play a crucial role in testing MG scenarios. Precise measurements of the $L$-$T$ relation across a wide mass and redshift range will be essential for constraining the parameter space of screened gravity models.

Looking ahead, several avenues can further improve and extend the present analysis. On the theoretical side, incorporating more realistic treatments of baryonic feedback, non-thermal pressure support, and environmental effects within hydrodynamical simulations will be essential for reducing modeling uncertainties. On the observational side, combining X-ray data with complementary probes such as weak lensing, Sunyaev-Zel'dovich measurements, and galaxy kinematics will help break remaining degeneracies. Finally, extending this framework to include redshift evolution and cross-correlations with large-scale structure surveys will provide a powerful pathway toward robustly testing gravity on cosmological scales.

\bigskip 
\bibliography{draft_ml}

\end{document}